\documentclass[fontsize=11pt, paper=letter]{scrartcl}
\usepackage[utf8]{inputenc}
\usepackage[english]{babel}
\usepackage{authblk}
\usepackage{siunitx}
\usepackage{amsmath}
\usepackage{amssymb}
\usepackage[width=\linewidth]{caption}
\usepackage{graphicx}
\usepackage{hyperref}
\usepackage[includeheadfoot, headsep=.1in, margin={.8in,.8in}]{geometry}

\title{Refractive index measurements of single, spherical cells using digital holographic microscopy}

\author[1]{Mirjam~Sch\"urmann}
\author[1]{Jana~Scholze}
\author[1]{Paul~M\"uller}
\author[2]{Chii~J.~Chan}
\author[1,3]{Andrew~E.~Ekpenyong}
\author[2,4]{Kevin~J.~Chalut}
\author[1,2]{Jochen~Guck}

\affil[1]{Biotechnology Center, Technische Universit\"at Dresden, Tatzberg 47/49, 01307 Dresden, Germany}
\affil[2]{Cavendish Laboratory, Department of Physics, University of Cambridge, JJ Thomson Ave, Cambridge~CB3~0HE,~UK}
\affil[3]{Department of Physics, Creighton University, 2500 California Plaza, Omaha, NE 68178, USA}
\affil[4]{Wellcome Trust/Medical Research Council Stem Cell Institute, Tennis Court Road, Cambridge CB2 1QR, UK}

\date{}

\begin{document}

\maketitle

\abstract{\textbf{Abstract.} In this chapter, we introduce digital holographic microscopy (DHM) as a marker-free method to determine the refractive
index of single, spherical cells in suspension. The refractive index is a conclusive measure in a biological context.
Cell conditions, such as differentiation or infection, are known to yield significant changes in the refractive index.
Furthermore, the refractive index of biological tissue determines the way it interacts with light. Besides the
biological relevance of this interaction in the retina, a lot of methods used in biology, including microscopy, rely on
light-tissue or light-cell interactions. Hence, determining the refractive index of cells using DHM is valuable in many
biological applications. This chapter covers the main topics which are important for the implementation of DHM:
setup, sample preparation and analysis. First, the optical setup is described in detail including notes and suggestions
for the implementation. Following that, a protocol for the sample and measurement preparation is explained. In the
analysis section, an algorithm for the determination of the quantitative phase map is described. Subsequently, all
intermediate steps for the calculation of the refractive index of suspended cells are presented, exploiting their
spherical shape. In the last section, a discussion of possible extensions to the setup, further measurement
configurations and additional analysis methods are given. Throughout this chapter, we describe a simple, robust, and
thus easily reproducible implementation of DHM. The different possibilities for extensions show the diverse fields of
application for this technique.}


\paragraph{Copyright.} Reproduced from \textit{Methods in Cell Biology} Volume 125, Chapter 9 (identical title and authors), Pages 143-159, Copyright (2015), \url{http://dx.doi.org/10.1016/bs.mcb.2014.10.016} with permission from Elsevier.

\section*{Introduction}

Digital holographic microscopy (DHM) is a quantitative phase microscopy technique which allows the measurement of the
phase shift introduced by an object in an interferometer. Other common imaging techniques based on interferometry, such
as phase contrast (PC) or differential interference contrast (DIC) microscopy, use optical path differences to image
phase variations in samples that are otherwise barely visible in bright-field microscopy. In contrast, DHM is used to
measure phase changes quantitatively. The measured phase contains two sample-specific characteristics, the thickness
and the refractive index (RI). 

There are different approaches to use the gained information in a biological context. The determined phase can be used
as a combined measure for thickness changes and RI changes of cells. For example, in the case of volume changes, the
thickness as well as the RI is altered. These two quantities cannot be separated just by taking the phase into account.
To study RI or thickness separately, either of the two parameters has to be known. In the course of this chapter, a method will be described on how to determine the RI of spherical cells. However, the principle stated here can also be adapted to the analysis of other samples of known shape. One of the three parameters phase, RI or thickness can then be used as output parameter of DHM measurements.

In several studies DHM has been successfully utilized to quantify cellular function. Taking the phase into account,
Pavillon et al. found distinctive features in DHM measurements of mouse cortical neurons they attribute to cell volume regulation as an early indication of cell death \cite{Pavillon2012}. Determining the RI from the measured phase values, Chalut et al. found lineage specific differences of the RI during the differentiation of a myeloid precursor cell line \cite{Chalut2012}. In a further example, a decrease in the RI of
Salmonella-infected macrophages has proven to be a reliable read-out for host-pathogen interactions \cite{Ekpenyong2012}. Besides using the RI as a measure of cellular function, the acquired information can also be integrated into other biophysical techniques that directly depend on the RI of the sample. One example is the measurement of optically induced stress in the context of optical traps. In any trap geometry where optical forces act on a cell, as in the case of an optical stretcher, the measured mechanical response depends on the RI of the cell and therefore needs to be evaluated separately \cite{Guck2005}. Another field of application is the interpretation of light scattering by tissues, aiming for example at marker-free sensitivity to tissue alteration in dysplasia. Here, the RI of the cells is one parameter influencing the analysis of the scattered light field \cite{Drezek2003, Perelman1998}.
Furthermore, studies on retinal tissue must deal with light-tissue interactions, where the optical properties of the tissue play a crucial role. Studies on photoreceptor nuclei revealed their ability to focus transmitted light \cite{Solovei2009, Blaszczak2014}. The ability to act like a lens is highly dependent on the RI
distribution inside these cells. All of the mentioned examples show that the RI of a cell and its distribution is a highly relevant quantity in certain areas of biology and biotechnology. 

Early measurements of the RI of living cells used RI matching of the measured object with the surrounding medium \cite{Barer1954}. In these measurements, the RI of cells in suspension is determined by adaption of the RI of the
medium to that of the cells. Index matching is achieved by gradually changing the RI of the medium until the contrast
in PC images between the cells and the medium is minimized. The RI of the medium corresponds to the RI of the cells at
the point of a reversal in sample contrast. Refractive index matching is usually performed with several cells inside the field of view leading to an average RI value for a cell population. However, this method does not allow for an accurate determination of RI on a single cell level and it is time consuming. Quantitative phase microscopy overcomes these technical constraints and allows measuring the RI of single cells.

There are different methods to measure the phase retardation of light transmitted through an object. A common concept, also used in DHM, is light interferometry which effectively measures the phase difference induced by a sample. The optical path difference at each position is the product of the thickness of the sample and the average RI of this path.
Hence, knowing the thickness of the sample at each position is a prerequisite for the determination of its RI. There
have been several attempts in the past to address this problem, for example by exchanging the medium surrounding the cells \cite{Rappaz2005} or measuring with two different wavelengths \cite{Rappaz2008b}. However, this involves rather elaborate sample preparation or setup construction. A different approach to separate the refractive index from the thickness of the sample is the combination of phase measurements with confocal microscopy. Confocal z-stacks from fluorescently labeled cells are used to determine the height at certain sample positions and the phase is measured at the same position in a subsequent measurement step \cite{Curl2005}. Nevertheless, due to the z-stacking and the related post processing, this technique is time-consuming.
Furthermore, the fluorescence labeling of the cells contradicts the maker-free nature of quantitative phase imaging. A simpler solution exists when the samples are transparent or semi-transparent spherical dielectric objects such as spherical cells. In this case the height at each position can be calculated from the determined radius of the cell in the field of view \cite{Kemper2007}. With this knowledge the average RI at each lateral position of the cell can be calculated. This approach will be elaborated further later in this chapter.

In the next section, a basic version of the DHM setup will be described in detail, followed by a protocol for
preparation of spherical cells in suspension. After a thorough guidance through the analysis method, possible
extensions or alternatives will be addressed in the discussion part.

\section{Setup}

The basic concept of digital holographic microscopy is to measure the phase change induced by a sample in an
interferometer. In one of the light paths, the so-called object beam, light is transmitted through the sample which
leads to a characteristic alteration of the wavefront. The second path serves as a reference for the measurement of the phase retardation and is accordingly termed reference beam (Figure 1). The resulting interference pattern is recorded by a complementary metal-oxide-semiconductuor (CMOS) camera.

\begin{figure}[ht]
\begin{center}
\includegraphics[width=.5\linewidth]{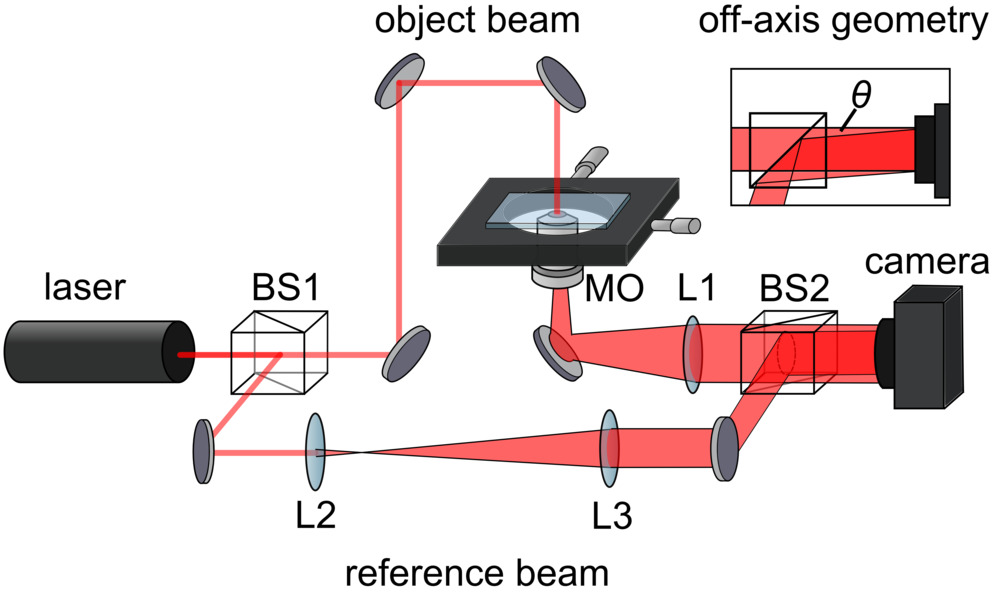}
\end{center}
\caption{Schematic drawing of the setup of the digital holographic microscope. All parts and their abbreviations are described in the text. The inset shows a top view of the off-axis geometry.}
\end{figure}

The light source is a HeNe-laser ($\lambda$ = 632.8 nm, HNL050L-EC, Thorlabs GmbH, Dachau, Germany) with a power of
about 0.5 mW. A non-polarizing beam-splitter cube (BS1) is used to split the beam with a ratio of about 50:50 into
object and reference beam. The cells in solution are mounted horizontally on a microscope slide. Allowing for lateral movements in the field of view, the microscope slide is placed on a two dimensional translation stage. The object beam is first steered over three mirrors so that it hits the sample vertically from the top (for a close up view of the wavefront retardation by the cell see Figure 2, which will be discussed in detail further below). Beneath the sample, a microscope objective (MO, 40x, NA 0.75, EC Plan-Neofluar air objective, Carl Zeiss Microscopy GmbH, G\"ottingen, Deutschland) is used to collect the retarded wavefront.
Mounting the MO on an axial translation stage (not shown for simplicity) enables focusing. After being focused in the MO the beam is divergent and has a spherical wavefront. This additional spherical phase pattern leads to subsequent difficulties in quantitative analysis. Hence, the divergent beam is collimated by a lens (L1) with a focal length of 200 mm.
This leads to the formation of a parallel wavefront and eliminates the spherical phase pattern \cite{Sanchez-Ortiga2011}. To check for proper alignment, a shearing interferometer (Shearing Interferometer, Thorlabs GmbH, Dachau, Germany) can be
used for testing the collimation of the beam while aligning the setup.
In the reference path a telescope built from two
lenses with focal lengths of 100 mm (L2) and 400 mm (L3) are used to adapt the beam diameter to the size of the camera chip. Again, the collimation of the reference beam is checked using a shearing interferometer. To be able to adjust the relative intensities of the two beam paths, a filter wheel (continuously variable ND Filter, Thorlabs GmbH, Dachau, Germany) is introduced in the reference path (not shown for clarity).
A second beam-splitter cube (BS2) is used to
overlap the two beams.Reference and object beam are built to have about the same length between BS1
and BS2.
The overlap of the two beams is implemented in a so-called off-axis geometry (see inset in Figure 1). This
means that the object and reference beam are not fully aligned on the same optical path behind BS2 but the two beam paths form a small angle  $\theta $ (not more than a few degrees), which is important for the analysis. 
Overlapping the two collimated beams leads to an interference pattern of parallel lines superimposed with an image of the sample from the object path (see Figure 3(A)). 
The interference pattern is recorded with a CMOS camera placed behind BS2. Since a laser is a coherent light source, it is important to use a camera without a glass cover in front of the chip to avoid additional detrimental interference artifacts.
The monochrome CMOS camera used here has 1280x1024 pixels with a pixel size of 5.2 \SI{}{\um} (MCE-B013-UW, Mightex Systems, Pleasanton, California, USA). The periodicity of the interference fringes depend on the angle  $\theta $  between the overlapping beams.
For reasons that will be discussed later in the analysis section the period of the interference pattern has to be small. However, a restriction for the period of the
interference pattern is given by the pixel size, i.e. the periodicity needs to be well resolved. For the particular
setup and camera described here, a period of about 4.4 pixels per line pair which translates to an angle  $\theta $  of about 1.6 deg works well. When imaging cells with DHM in this configuration, it is possible to capture more than one cell in one image. However, for the analysis of the cells it is important that they are well separated. For each measurement a background image is recorded additionally by moving the cell slightly out of the field of view with the translation stage.
The recorded images are analyzed to obtain a two dimensional phase map of the cell.

\section{Measurement Preparation}

Before starting the measurement, the RI of the medium and the magnification of the microscope have to be determined. A successful measurement of the RI of spherical cells in solution requires a specific sample preparation.  The cells need to be attached to a cover slip in order to stay in focus.
Additionally, the cells need to be spherical during the
measurement so that the height can be determined correctly (see Figure 2). The following protocol serves as a basic
guideline for measurement preparations.

\begin{figure}[ht]
\begin{center}
\includegraphics[width=.34\linewidth]{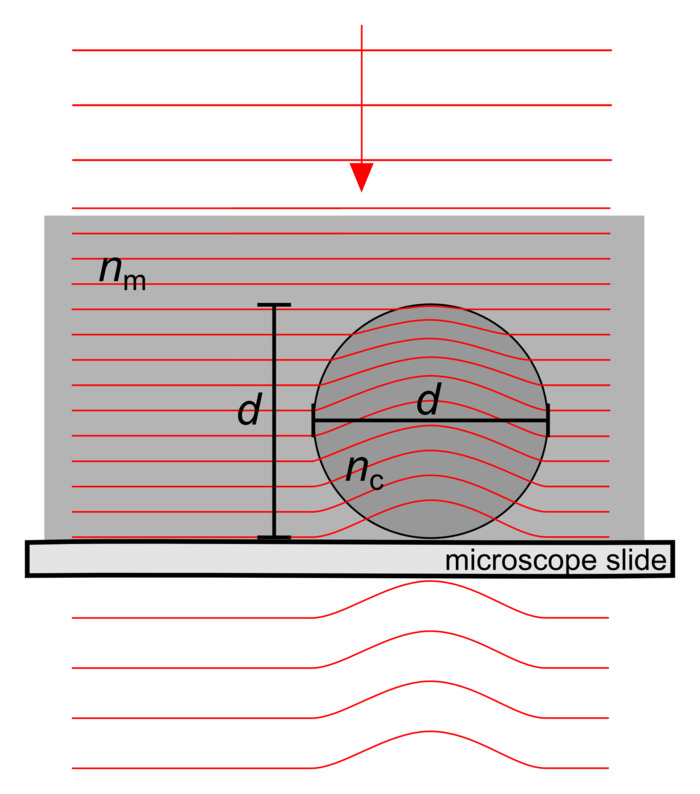}
\end{center}
\caption{Schematic drawing of a spherical cell on a microscope slide depicting the change in wavefront as it traverses the cell. Here,  $d$  is the diameter of the cell,  $n_m$  is the RI of the medium and  $n_c$  the RI of the cell.}
\end{figure}

\subsection{Materials}
\begin{itemize}
\item Slide with micrometer scale (graticule)
\item Refractometer (e.g. Abbe-Refractometer 2WAJ, Arcarda GmbH, Reichelsheim, Germany)
\item Glass cover slip
\item Poly-L-Lysine (PLL) 0.01\% solution or Poly-D-Lysine (PDL)
\item Phosphate-buffered saline (PBS)
\item Cells and medium
\end{itemize}

\subsection{Method}
\begin{enumerate}
\item Scale Determination

The magnification of the system can be determined with the help of a graticule. Even though the magnification
usually does not change significantly over time, this can be repeated on a regular basis before starting a measurement
to consider instabilities due to e.g. temperature drifts.

\item RI of Medium

Before every measurement, the RI of the medium needs to be measured using a refractometer.

\item  Cell Preparation
\begin{enumerate}
\item Take 2--3 mL of cells out of the culture dish
\item Centrifuge and resuspend in a medium of choice for a final concentration of approximately 3 x
10\textsuperscript{5} cells/mL
\item Place 100 \SI{}{\uL} of PLL on the cover slip and let it settle for 10 min
\item Remove PLL and wash twice with PBS 
\item Remove PBS from cover slip and add 100--120 \SI{}{\uL} of cell suspension
\item Wait for 10 min for the cells to attach to the bottom
\item Carefully remove non-attached cells by pipetting off most of the cell suspension (be careful not to ``dry'' the
cells) and add approximately the same amount of fresh medium 
\end{enumerate}
\end{enumerate}

\subsection{Comments}
\begin{itemize}
\item Coating the microscope slides with PLL leads to a charge based attachment of cells.
For every cell type one has to make sure that the above protocol fulfills the following requirements. The cells have to attach to the microscope slide and must retain their spherical shape without extensively spreading on the microscope slide during the measurement period (about 15 min).
Spherical shape can be controlled for example by using confocal fluorescence microscopy with a
well-calibrated microscope.
\item The difference of the RI of the medium to the average RI of the cells should not be less than about 0.001,
otherwise the contrast will be too low for a successful analysis. Furthermore, the difference should not be more than about 0.1, otherwise the phase jump at the object-medium interface might not be resolved. These requirements are usually fulfilled for cells in water-based solutions. 
The upper limit can be reached when measuring artificial objects such as polystyrene beads that, in addition to their high refractive index compared to water, exhibit very sharp boundaries.
\item For the measurement of the RI of the medium, an Abbe refractometer with an accuracy of 0.001 is needed. 
\end{itemize}

\section{Data Analysis}
The goal of the analysis is to determine the RI of an individual cell. In essence, this is done by converting the phase shift information captured in the interference image into a quantitative RI. We implemented the different analysis steps of the algorithm in LabVIEW. A detailed description of the image analysis is presented below (code available upon request).

\begin{figure}[hp]
\begin{center}
\includegraphics[width=.9\linewidth]{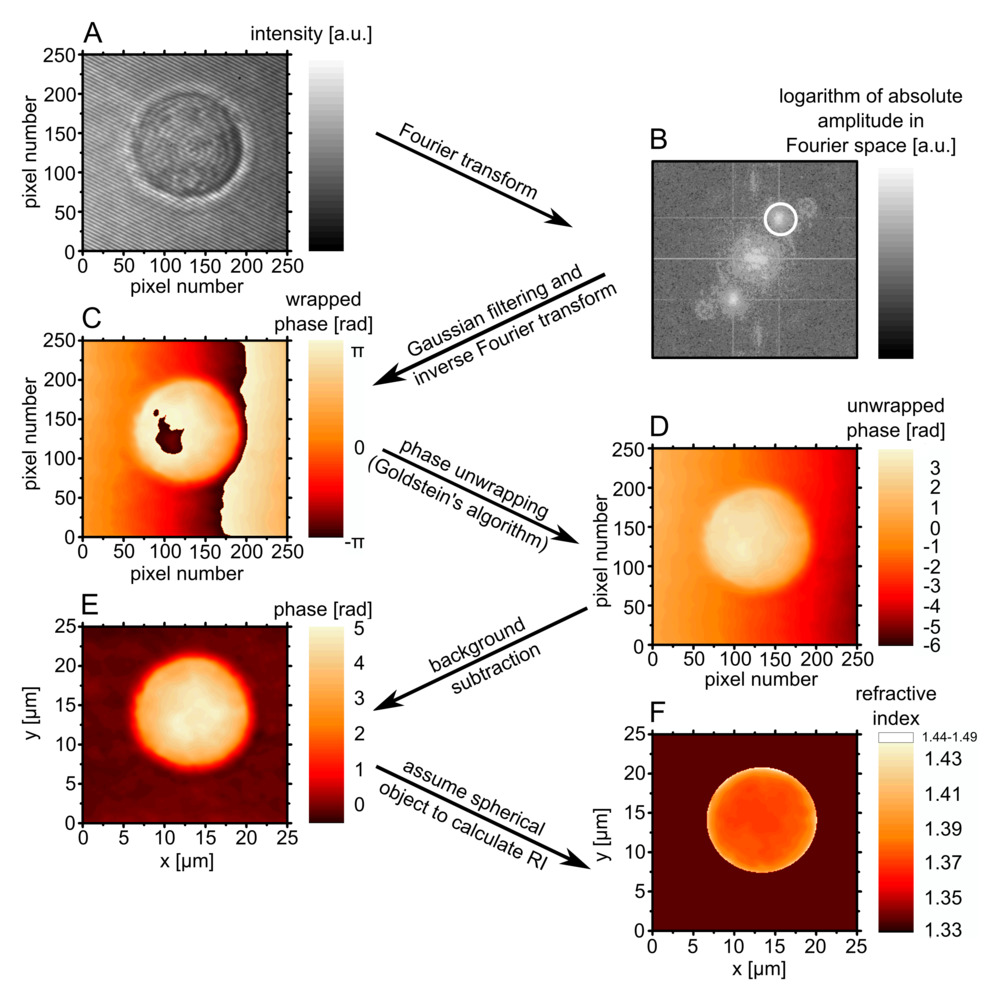}
\end{center}
\caption{Analysis steps to determine the refractive index (RI) of an HL60 cell from a digital holographic image. (A)
Raw intensity image of CMOS camera. (B) Logarithm of absolute value of Fourier transform of the intensity image. 
The zero frequency component lies in the center of the shown Fourier space image. The white circle indicates the position of the Gaussian filter. (C) Wrapped phase image. (D) Unwrapped phase image.
(E) Unwrapped phase with background correction. (F) RI distribution.}
\end{figure}

\subsection{Determination of the Phase Distribution}

The intensity image or hologram (Figure 3(A)) obtained from the DHM is an interference image resulting from the overlap
of the reference wavefront ($R$) and the object wavefront ($O$). The critical phase information is encoded within the
interference pattern. To extract the phase information, we need to consider the intensity  $I$  of the overlapping
beams: 
\begin{equation}
I=\left|R^2\right|+\left|O^2\right|+R^{\ast }O+O^{\ast }R
\end{equation}
where  $R^{\ast }$ and  $O^{\ast }$ are the complex conjugates of the wavefront of the reference beam and the object
beam respectively. The first two terms of equation (1) contain information about the amplitude of the individual waves.
The last two terms result from the interference of object  $O$  and reference  $R$  wave. They include the information
about phase and amplitude of the object and reference wave. To be able to separate the different terms of the recorded
intensity, a Fourier transform is applied.
Due to the interference pattern in the off-axis geometry the terms that contain the phase information are shifted away from the origin in Fourier space \cite{Cuche2000}.
In contrast, the first two intensity terms in equation (1) stay at the origin in Fourier space. For a better
visualization of the result of the Fourier transform, the absolute value is plotted on a logarithmic scale in Figure
3(B). Here, the first two terms of equation (1) contribute to the peak in the center of Fourier space whereas the last
two terms each give rise to one of the side peaks. To improve the image contrast in Fourier space, the central peak is
suppressed by subtracting the mean intensity value from each pixel in the raw intensity image prior to the Fourier
transform \cite{Cuche2000}. The phase information of interest is equally encoded in both side peaks and only one of
the peaks is required for phase retrieval. For this, a filter is applied in Fourier space that removes all other
contributions besides the peak chosen (white circle in Figure 3(B)). We apply a Gaussian filter to reduce ringing
artefacts in the reconstructed phase. After filtering, the frequencies are shifted to the center of Fourier space.
Subsequently an inverse Fourier transform is applied yielding the complex field \textit{F} from which the phase can be
calculated \cite{Cuche2000}. 
\begin{equation}
\Delta \varphi =\arctan \left(\frac{\Im (F)}{\Re (F)}\right)
\end{equation}
The choice of the filter width is important. If the width is too wide, frequencies of the central peak may affect the
reconstructed image. If the filter is too small, crucial higher frequencies might be omitted which lowers the
resolution of the phase distribution. The positions of the side peaks are dependent on the period of the interference
pattern. As discussed earlier, this period is given by the angle  $\theta $  of the off-axis geometry. Therefore, the
distance from the side peaks to the center of Fourier space can be adjusted by varying the off-axis angle in the setup.
Good filtering requires that the side peaks are well separated from the peak at the center of Fourier space. This can
be achieved by choosing  $\theta $  such that the interference pattern has a small period.

Note that the phase calculated from equation (2) only contains values between -$\pi $ and $\pi $. However, the physical
phase shift may contain values outside of this range (modulo 2$\pi $). In that case, the computed phase image contains
sudden phase jumps of 2$\pi $ (Figure 3(C)) and is termed wrapped phase image. Thus, a crucial part of the analysis is
to unwrap the phase. This can be done with the help of Goldstein's algorithm \cite{Ghiglia1998}. The underlying
principle of the algorithm is to scan for 2$\pi $ phase jumps within the wrapped data and to add an additional phase of
2$\pi $ to the part of the image with the lower phase. A representative image of the unwrapped phase is shown in Figure
3(D).

As discussed earlier, the distance of the side peaks to the center of Fourier space is determined by the angle  $\theta
$  of the off-axis geometry. If the Gaussian filtered region in Fourier space is not shifted to the exact center before
Fourier inversion, the phase image will be superimposed by a linear ramp whose slope is dependent on the angle  $\theta
$. Since the shift in Fourier space is discrete, the ramp cannot be fully removed, and a correction of the unwrapped
image is necessary. There are two ways to accomplish this correction. One possibility is to take a background image as
described in the setup section and to subtract the background phase from the measurement phase. Before subtraction, the
background image has to be processed exactly the same way as the measurement image. A second possibility is to apply a
ramp correction. A linear fit in the x- and y-direction to regions of the phase image that are known to be background
is used to evaluate the slope and the offset of the ramp. This computer generated background image is then subtracted
from the unwrapped phase image. The ramp and offset correction can also be applied in series with the background
correction in case of marginal drifts between the measurement and background images. Applying both background
correction and ramp correction gives the best results. If the setup or the sample do not allow recording an additional
background image, a ramp correction alone can be sufficient. After the background correction, one obtains a two
dimensional phase distribution of the studied cell, as shown in Figure 3(E).

\subsection{Determination of the refractive index}

The last step is to calculate the average RI from the phase distribution. When light passes through an object with the
optical path length  $s=h{\cdot}n$, where  $h$  is the thickness and  $n$  the RI of the object, it experiences a phase
shift  ${\Delta}\varphi $  compared to light that does not pass this object
\begin{equation}
{\Delta}\varphi =\frac{2\pi }{\lambda }{\Delta}s
\end{equation}
where  ${\Delta}s$  is the optical path difference between these two light paths and  $\lambda $  is the vacuum
wavelength of the light.

In DHM measurements the phase shift due to the cell  $\Delta \varphi _c\left(x,y\right)$  is the quantity of interest.
To measure this phase shift, the cell with an axially averaged RI distribution  $n_c(x,y)$, an effective axial
thickness profile  $h_c\left(x,y\right)$  and a diameter  $d$ (see Figure 2) is suspended in a medium with a known RI 
$n_m$  and is introduced into the object path of the setup. An additional phase shift is present in the resulting phase
image due to an optical path difference  ${\Delta}s_0(x,y)$  between object and reference path which results e.g. from
the optical components, the medium above the cell in the object path or slight differences in the lengths of the
reference and object path. The path difference  ${\Delta}s_0(x,y)$  is present in every measurement and can be
considered as a setup-specific background. 

The measured optical path difference between object and reference path includes both the contribution of the cell as
well as the background and can be written as:
\begin{equation}
\frac{{\Delta}\varphi (x,y)\lambda }{2\pi
}=h_c\left(x,y\right)n_c\left(x,y\right)+\left(d-h_c\left(x,y\right)\right)n_m+{\Delta}s_0\left(x,y\right)
\end{equation}
For reasons of simplicity the dependence of  ${\Delta}\varphi $, $h_c$,  $n_c$  and  ${\Delta}s_0$  on the coordinated 
$x$  and  $y$  will be omitted in the following. Equation (4) can be rewritten as: 
\begin{equation}
\frac{{\Delta}\mathit{\varphi \lambda }}{2\pi
}=h_c\underbrace{(n_c-n_m)}_{{\Delta}s_c}+\underbrace{dn_m+{\Delta}s_0}_{{\Delta}s_{\mathit{BG}}}
\end{equation}
Here, the first part on the right side of the equation is the cell-specific optical path difference  ${\Delta}s_c$. The
second part, termed  ${\Delta}s_{\mathit{BG}}$, is the total background which is subtracted according to the previous
section. Inserting  $h_c=0$  in equation (5) yields the phase of the background image, i.e. with medium but without
sample in the light path $:$
\begin{equation}
\frac{{\Delta}\varphi _{\mathit{BG}}\lambda }{2\pi }=dn_m+{\Delta}s_0
\end{equation}
Hence, subtracting the phase shift of the background image (equation (6)) from the phase shift measured with the cell in
the field of view (equation (5)) yields the cell-specific phase shift  $\Delta \varphi _c$. Rewriting this difference
results in an expression for the axially average refractive index of the cell at each lateral position:
\begin{equation}
n_c=n_m+\frac{{\Delta}\varphi _c\lambda }{2\pi h_c}
\end{equation}
The only unknown parameter on the right-hand side of equation (7) is the thickness of the cell  $h_c$. Assuming a
spherical cell, the cell's thickness can be calculated from the radius at every point of the cell. The radius is
obtained by fitting a circle to the edge of the cell in the phase image. The edges can be determined with an edge
detection algorithm, e.g. Canny edge detection (included in LabVIEW). Knowing the thickness of the cell at each
position, the integral RI for each pixel within the cell is calculated. The points near the edge are excluded since the
thickness at these points approaches zero and the RI becomes ill defined. The RI of the cell is determined as a
weighted average, where all individual integral refractive index values within the perimeter of the cell are weighted
with  $h_c$. This emphasizes RI values at positions with a longer optical path length. In order to capture the average
RI value of a whole cell population reliably one has to acquire a large enough sample size to obtain a normal
distribution of the cells' RI values. 

\section{Discussion}
The protocol described so far provides the reader with the basic know-how for building a simple DHM setup that is able
to deliver the phase and RI of individual spherical cells. However, there are various possibilities to extend or
replace parts of the setup and analysis method which will be discussed in the following sections. 

\subsection{Setup Alterations}

One approach to further enhance digital holographic microscopy is the improvement of the quality of the acquired
intensity images and consequently the gathered phase maps. This can be especially of interest if the goal is a detailed
analysis of phase or RI changes within a single cell. One possibility in this direction is noise reduction. In the
setup described, the laser is used as the light source, which being a source of coherent light, may introduce
additional interference patterns due to back reflections at glass interfaces or other obstacles in the beam path. A
reduction of this unwanted interference can be achieved using a light source with a shorter coherence length such as a
light emitting diode (LED) \cite{Kemper2008} or by reducing the
spatial coherence of a laser with a rotating ground glass in the beam path \cite{Dubois2004}. In comparison to setups that use lasers with a long coherence length, the path length of the
reference and the object beam need to be adjusted more precisely for light sources that have shorter coherence lengths.
At short coherence length, interference between object and reference beam, which can be used for analysis, requires
that the two beam paths have exactly the same length, which therefore requires a more accurate alignment. 

There is also the possibility to integrate a system for quantitative phase measurements into commercially available
microscopes, rather than building the setup from scratch. The integration of a digital holographic microscope into a
commercial microscope is one option \cite{Kemper2006}. Furthermore, there is the option to buy a digital
holographic microscope as a complete system (e.g. Lync\'ee Tec SA, Lausanne, Switzerland). Alternatively, a camera
upgraded with a diffraction grating (SID4-BIO, Phasics S.A., Palaiseau, France) can be used, which uses the principle
of quadri-wave lateral shearing interferometry to determine the quantitative phase shift \cite{Bon2009}. This technique does not require any changes to the light path of the microscope, and the software
acquires the quantitative phase readily; therefore the analysis only requires the determination of the RI from the
measured phase. 

\subsection{Assumption of Spherical Shape for Refractive Index Determination}

The prerequisite for the analysis method described so far is the assumption that the studied sample must be spherical in
shape. In the protocol described here, there is the possibility that, with time, cells spread on the coverslip that
they are attached to. Therefore, the validity of the assumption of spherical cells needs to be checked when
establishing this technique for a new cell type. Cells often show spherical shapes when they are not attached to any
surface but are floating freely in solution. To still be able to control the position of cells in solution for in-focus
imaging, an optical trap can be used \cite{Kemper2010}. However, also in this case cells will never resemble a
perfect sphere. Furthermore, the recorded phase images contain only the projected phase which limits the analysis in
two dimensions. An approach which overcomes these limitations altogether is tomography. Imaging the cell from different
angles allows a complete volume reconstruction, such that the thickness can be directly determined from the data.
Furthermore, measuring from multiple angles allows computing a full three-dimensional RI map of the cell, which permits
the unambiguous assignment of the RI of smaller organelles. One possibility of RI tomography involves rotating the
sample with a micropipette \cite{Bergoend2010}. Another approach leaves the sample
stationary while changing the angle of incidence of the light in the object path by tilting a mirror \cite{Choi2007}. Sung et al. showed that it is possible to acquire three-dimensional phase maps in a microfluidic system \cite{Sung2014}.
They combine DHM with the synthetic aperture approach to determine three dimensional refractive index maps of single
cells.

\subsection{Analysis Alterations}

While we described our analysis method of choice above, there are numerous alternative or enhanced methods. For example
it is possible to numerically reconstruct the wave front using a digital reference wave. The digital reference wave
resembles the plane wave used during the acquisition of the hologram. In contrast to classical holography, where a
plane wave is needed for an optical reconstruction of the sample, in this method a numerical reconstruction is achieved
by multiplications of the hologram with a digital reference wave \cite{Cuche1999}. A further analysis approach allows to numerically refocus the gathered data which is achieved by
changing the reconstruction distance \cite{Cuche1999}. This makes focusing possible without mechanically moving the
specimen, which is very valuable for imaging dynamic processes. An even further extension is the use of numerical
autofocusing of the gathered data. Here, the optimal numerical focus is obtained by maximizing the sharpness of the
numerically focused image \cite{Langehanenberg2008}. The advantage of this method is that
only one image needs to be taken and that the autofocusing can be performed computationally in a subsequent step after
image acquisition. Yourassowsky and Dubois introduced a microfluidic channel in their setup to image plankton
organisms of different sizes and in different focus planes \cite{Yourassowsky2014}. They employed numerical focusing to overcome the problem of
different focal planes. Other possibilities of post-processing are methods to improve the image quality. This concept
was implemented by Colomb et al. \cite{Colomb2006}. They were able to overcome aberrations in the imaging system by implementing
numerical post-processing. A crucial part of the analysis is the phase unwrapping process. Several alternatives to
Goldstein's algorithm exist. One of these possibilities relies on the fact that changing the above mentioned numerical
reconstruction distance leads to a spatial shift of the  $2\pi $  phase jumps in the reconstructed field \cite{Khmaladze2010}. By computing phase images with multiple reconstruction distances an unwrapped phase image can
be obtained. 

While the above mentioned alterations correspond to alternatives for the phase extraction, there are also different
parameters that can be measured and measurement modalities that can be used in quantitative phase microscopy. For cells
that have a nearly uniform RI distribution, e.g. red blood cells, the thickness profile can be determined from the
phase \cite{Popescu2005}. Rappaz et. al compared the volume measurements of red blood cells with DHM to other
techniques including volume determination from confocal stacks and measurements performed with an impedance volume
analyzer \cite{Rappaz2008a}. They found that the determined volume of red blood cells with confocal
microscopy and DHM gives comparable results. DHM requires the acquisition of only one image to determine the volume of
a red blood cell. In comparison, in confocal microscopy a complete z-stack needs to be acquired to fulfill the same
task. However, one has to be aware of the fact that the analysis of red blood cells is not representative for the
general analysis of cells. The determination of the volume depends on the homogeneity of the refractive index of red
bloods cell. This assumption is valid, because red blood cells lack internal structures and are filled with hemoglobin
at constant concentration throughout. The determination of the volume using a single DHM image is not possible in the
case of large RI variations and an unknown RI distribution. Another application of phase imaging, is suitable for cells
that are highly dynamic, e.g. contracting cardiomyocytes, here temporal measurements can be used to characterize their
cell dynamics \cite{Shaked2010}. A further example involves the determination of the dry mass
of cells \cite{Zangle2013}. All these numerous alternatives and add-ons show the
diverse ways in which data acquisition and analysis can be adjusted to meet specific measurement goals.

\section{Summary}

This chapter gives an overview of all the steps needed for the implementation of digital holographic microscopy. The
setup, sample preparation and analysis routines are described in detail and allow for an easy but robust determination
of the phase and RI map of spherical cells. We also discuss further extensions to the proposed setup and present a
brief overview of the diverse biological applications. The methods described in this chapter form the basis for
versatile applications in biological studies.

\bibliographystyle{ieeetr}
\bibliography{chapter}

\end{document}